\documentclass{aa}
\usepackage{graphicx,epsf,epsfig}

\usepackage{txfonts}
\begin{document}
   \title{The X-ray properties of the magnetic Cataclysmic Variable 
UU\,Col\thanks{Based on observations obtained with XMM-Newton, an ESA
science mission with instruments and contributions directly funded by ESA 
Member States and NASA.}}


   \author{D. de Martino
          \inst{1}
          \and
          G. Matt\inst{2}\and K. Mukai\inst{3}\and
J.-M. Bonnet-Bidaud\inst{4}, V. Burwitz\inst{5}, B.T. G\"ansicke\inst{6},
F. Haberl\inst{5} \and M. Mouchet\inst{7}
          }

\offprints{D. de Martino}

   \institute{INAF--Osservatorio Astronomico di Capodimonte, Via
Moiariello 16, I-80131 Napoli, Italy\\
              \email{demartino@na.astro.it}
         \and
Dipartimento di Fisica, Universita' degli Studi Roma Tre, Via della
Vasca Navale 84, I-00146 Roma, Italy \\
\email{matt@fis.uniroma3.it}
\and
Laboratory for High Energy Astrophysics, NASA/GSFC, Code 662, Greenbelt,
MD 20771, USA\\
 \email{mukai@milkyway.gsfc.nasa.gov}
\and
Service d'Astrophysique, DSM/DAPNIA/SAp, CE Saclay, F-91191 Gif sur Yvette
Cedex, France\\
\email{bonnetbidaud@cea.fr}
\and
Max-Planck-Institut f\"ur Extraterrestrische Physik,
Giessenbachstra{\ss}e, Postfach 1312, 85741 Garching, Germany \\
\email{burwitz@mpe.mpg.de, fwh@mpe.mpg.de}
\and 
Department of Physics, University of Warwick, Coventry CV4 7AL,
UK\\ 
\email{boris.gaensicke@warwick.ac.uk}
\and
APC, UMR 7164, University Denis Diderot, 2 place Jussieu, F-75005 and 
LUTH, Observatoire de Paris, F-92195 Meudon Cedex, France\\
\email{martine.mouchet@obspm.fr} 
             }

   \date{Received February 24, 2006; accepted March 11, 2006}

\abstract{}
{XMM-Newton observations aimed at determining for the first time 
the broad-band X-ray  properties of the  faint  high galactic latitude  
Intermediate Polar  UU\,Col are presented.} 
{We performed X-ray timing analysis in different energy ranges of the EPIC 
cameras which reveals the dominance  of the 863\,s white dwarf rotational  
period. The spin pulse is strongly energy dependent. Weak 
variabilities at the beat 935\,s and at 
the 3.5\,hr orbital periods are also observed, but the orbital modulation 
is detected only below 0.5\,keV. Simultaneous UV and optical photometry 
shows that the spin pulse is anti-phased with respect to the hard 
X-rays. Analysis of the EPIC and RGS spectra reveals the complexity 
of the X-ray emission, being composed of a soft 50\,eV black--body 
component 
and  two optically thin emission components at 0.2\,keV and 11\,keV 
strongly absorbed 
by dense material with an equivalent hydrogen column density of 
$\rm 10^{23}\,cm^{-2}$ partially (50$\%$) covering the X-ray source.} 
{The complex X-ray and UV/optical temporal behaviour indicates that 
accretion occurs predominantly ($\sim80\%$) via a disc with a  
partial contribution ($\sim20\%$) directly from the stream. The main 
accreting pole dominates at high energies whilst the secondary  
pole mainly contributes in the soft X-rays and at lower energies. 
The bolometric flux 
ratio of the soft-to-hard X-ray  emissions is found to 
be consistent with the prediction of the standard accretion shock model. 
We find the white dwarf in UU\,Col accretes at a low rate and 
possesses a low magnetic moment. 
It is therefore unlikely that 
UU\,Col will evolve into a moderate field strength Polar, which leaves the
soft X-ray Intermediate Polars a still enigmatic small group of magnetic 
Cataclysmic 
Variables.} 
{}


   \maketitle
%

\section{Introduction}

The Intermediate Polars (IPs), a subclass of magnetic Cataclysmic
Variables (mCVs), are the most luminous and hardest X-ray
sources  among accreting white dwarfs (WDs). They contain a magnetized 
asynchronously rotating WD ($\rm P_{\omega} << P_{\Omega}$) accreting from 
a late type, Roche-lobe filling star, usually via a truncated disc.
Accretion falls Ãquasi-radially towards the magnetic poles, forming a
standing shock below which material cools via hard X-rays and cyclotron 
emission. They differ from the Polars, the other subclass of mCVs, 
which instead are synchronous, strong emitters of soft X-rays as well as 
of optical/near-IR polarized radiation.  Due  to the 
intense (10-230\,MG) WD magnetic field, the Polars do  not possess an 
accretion disc. The soft X-ray emission is believed to 
arise from reprocessing of hard X-rays and cyclotron radiation in the WD
atmosphere. The ratio of hard to soft X-ray fluxes
strongly depends on the magnetic field strength, reflecting the
interplay between thermal bremsstrahlung and cyclotron cooling, with the 
latter dominating in high field systems (Woelk \& 
Beuermann \cite{WoelkBeuermann}). 
The lack of soft X-ray emission in the majority of IPs was  
explained by the higher accretion rates, the high intrinsic 
absorption and the larger accreting areas with respect to the Polars 
shifting the peak of reprocessed emission to the EUV/UV range. 
The absence of polarized radiation in the optical and near-IR in the 
majority of these systems also   
led to suggest that they  possess lower magnetic field strength WDs than 
the Polars.
Whether IPs will evolve into Polars or they represent a distinct class is 
still a debated evolutionary 
issue  (Cumming \cite{Cumming}, Norton et al. \cite{norton04}).
A small group of four IPs (PQ\,Gem, V405 Aur, UU\,Col and 1RXS 
J062518.2+733433) was recognized to also possess a 
soft X-ray emission  component (Haberl \& Motch \cite{Haberl95}, 
Burwitz et al. \cite{Burwitzetal},  Staude et al. \cite{Staude03}), 
similar to that observed in Polars with two optically bright systems 
(PQ\,Gem and V405 Aur) also showing optical polarized 
radiation. Their similarity with low field Polars led to the suggestion 
that these IPs (also called "soft IPs") 
could be their true progenitors. Recently, two hard 
X-ray IPs were discovered to possess a soft X-ray component 
but highly  absorbed and hotter than that of the soft IPs and Polars 
(Haberl et al. \cite{Haberl02}, de Martino et al. \cite{demartino04}), 
raising the  question on  whether a soft X-ray component is indeed 
present in all IPs. 

Notwithstanding this, since its discovery from the ROSAT All Sky Survey 
(RASS), the high latitude X-ray source 
1RXS\,J0512.2-3241=UU\,Col (henceforth UU\,Col)  identified as 
a soft X-ray IP (Burwitz et al. \cite{Burwitzetal}), remained 
mostly unnoticed. A follow-up X-ray ROSAT HRI pointed 
observation (Burwitz \& Reinsch \cite{Burwitz01}) confirmed the 
optical photometric periods  of 3.45\,hr and 863.5\,s interpreted as the 
binary period  and the rotational period of the accreting WD respectively. 
However,  a knowledge of its broad band X-ray properties had to await the 
advent of more sensitive X-ray facilities.

\noindent In this work we report on the first {\em XMM-Newton} 
observation of UU\,Col aimed at determining the variability 
characteristics in both X-ray and UV/optical domains as well as the 
X-ray spectral properties of this poorly studied mCV and at inferring 
the accretion and WD parameters to understand its evolutionary state.

   \begin{table*}[t!]
      \caption{Summary of the {\em XMM-Newton} observation of UU\,Col.}
         \label{obslog}
     \centering
\begin{tabular}{l  c c r c c}
            \hline \hline
            \noalign{\smallskip}
Instrument     &  Date & UT(start) & Duration (s) & Net Count Rate 
(cts\,s$^{-1}$)\\
            \noalign{\smallskip}
            \hline
            \noalign{\smallskip}
 EPIC PN &  2004 Aug 21 & 22:00 & 26037 & 0.50\\
 EPIC MOS & & 21:38 & 27666 & 0.18 \\
 RGS    & & 21:37 & 27918 & 0.018  \\
 OM B & & 21:46 & 2441 & 3.10\\
      & & 22:32 & 2441 & \\
      & & 23:18 & 2440 & \\
      & 2004 Aug 22 & 00:04 & 2440 & \\
      & & 00:50 & 2439 & \\
 OM UVM2 & & 01:36  & 2439 & 0.47 \\
         & & 02:22 & 2440 & \\
         & & 03:08 & 2440 & \\
         & & 03:54 & 2439 & \\
         & & 4:40 & 2439  & \\
                 \noalign{\smallskip}
            \hline
\end{tabular}
\end{table*}


\section{The XMM-Newton observation}
  
UU\,Col was observed with the {\em XMM-Newton\/} satellite (Jansen et
al. \cite{jansenetal}) on 2004 August 21
(obsid:0201290201) with the EPIC-PN (Str\"uder et
al. \cite{strudeetal}) and MOS (Turner et al. \cite{turneretal}) cameras
operated in full frame mode   with the thin filter for  a net
exposure time of
26.0\,ks and 27.7\,ks respectively. UU\,Col was also observed with the
Reflection Grating Spectrographs (RGS1 and RGS2) (den Herder et
al. \cite{denherderetal}) in
spectroscopy mode with an exposure time of
27.9\,ks and with the Optical Monitor (OM) instrument (Mason et
al. \cite{masonetal}) with the UVM2 and B filters
covering the ranges 2000--2800\,\AA \, and
3900--4900\,\AA \, in imaging fast mode for a total exposure time of
12.3\,ks in each filter. A summary of
the observations is reported in Table~\ref{obslog}.

   \begin{figure}
   \centering
\includegraphics[height=9.cm,width=9.cm,angle=0]{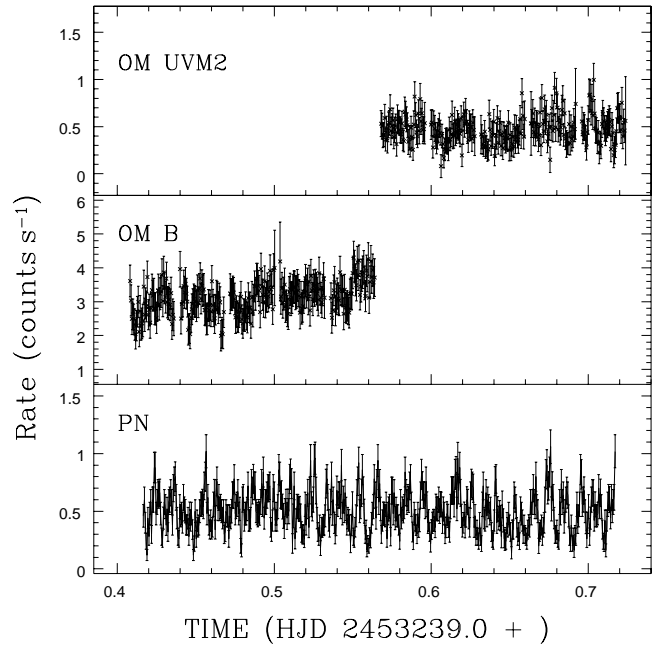}
\caption{{\em From bottom to top:} The light curves in the EPIC PN 
0.2--15\,keV range, in the B band and in the UV 2000--2800\,$\AA$ range
binned in 50\,s. }\label{fig1}
    \end{figure}

The standard processing pipeline data products were used. 
The EPIC light curves and spectra and the RGS spectra were extracted with
the SAS 6.5 package retrieved  from the  ESA-VILSPA Science Center.
Due to the proximity of UU\,Col to
the CCD No.1 border of the EPIC PN camera, the light curves and spectra  
were extracted
from a circular region with a radius of 17.5$^{"}$ centred on the source, 
while for the EPIC MOS cameras a larger extraction radius 40$^{"}$ was 
used. Background light curves and spectra were extracted from  offset
circular regions with same radii as for the target on the same CCD chip.  
Single and double pixel events with a
zero quality flag were selected for the EPIC-PN data, while for EPIC-MOS
cameras up to quadruple pixel events were  used.

The EPIC-PN and, at a less extent, the MOS data, are  affected by
only moderate background activity (up to 0.16\,cts\,s$^{-1}$) not 
influencing 
the light curve (see Fig.~\ref{fig1}). However, 
for the spectral analysis, we conservatively windowed the data in order 
to exclude epochs when background count rate exceeds 0.13\,cts\,s$^{-1}$ 
in the EPIC-PN camera, implying a $\sim$20$\%$ screening of the data. 
The extracted  EPIC-PN and MOS average 
spectra were then rebinned to have 
 a minimum of 20 counts in each bin, while phase--resolved spectra were 
rebinned with a minimum of 25 counts per bin to allow the use of the 
$\chi^2$ statistics.
Ancillary response and redistribution matrix files were created using SAS
tasks {\em arfgen} and {\em rmfgen} respectively.
  
The RGS pipeline was run using the SAS task {\em rgsproc}.
RGS1 and RGS2 first order spectra have been rebinned to have 
a minimum of 20 counts per bin. 

OM background subtracted light curves 
 produced  by the standard processing pipeline were used for timing 
analysis (see Fig.~\ref{fig1}).
  Average net count rates are 3.106\,cts\,s$^{-1}$ 
(B filter) and  0.471\,cts\,s$^{-1}$ (UVM2 filter), corresponding to
the instrumental magnitudes: B=18.0$\pm$0.6\,mag and
UVM2=16.6$\pm$0.7\,mag. Using Vega magnitude to flux conversion, these
correspond to a flux of $\rm 3.9\times 
10^{-16}\,erg\,cm^{-2}\,s^{-1}\,\AA^{-1}$
in the 3900--4900\,\AA \, band and of $\rm 1.02\times
10^{-15}\,erg\,cm^{-2}\,s^{-1}\,\AA^{-1}$
in the 2000--2800\,\AA \, band. The B band level is
consistent to that observed in 1996 by Burwitz et al. 
(\cite{Burwitzetal}).

   \begin{figure}[h!]
   \centering
\includegraphics[height=9.cm, width=8.cm,angle=-90]{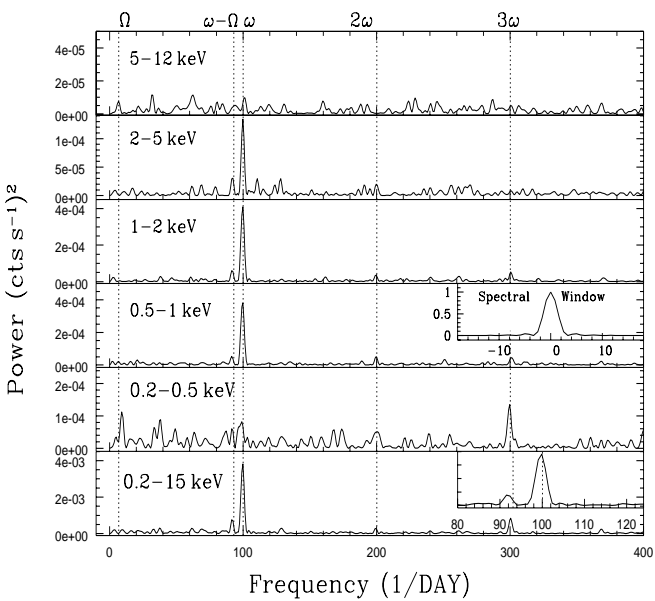}
\caption{EPIC-PN power spectra in selected energy ranges. From bottom 
to top: 0.2--15\,keV, 0.2--0.5\,keV, 0.5--1\,keV, 1--2\,keV, 2--5\,keV and
5--12\,keV. The spin ($\omega$) and harmonics, the beat ($\omega - 
\Omega$) and the orbital ($\Omega$) frequencies 
as derived from optical 
data (Burwitz et al. \cite{Burwitzetal}) are marked with vertical
dotted lines. Inserted panels show an enlargment around the spin 
frequency (lower panel) and the spectral window of the data.}\label{fig2}
    \end{figure}

   \begin{figure*}[t!]
   \centering
\mbox{\epsfxsize=8cm\epsfysize=8cm\epsfbox{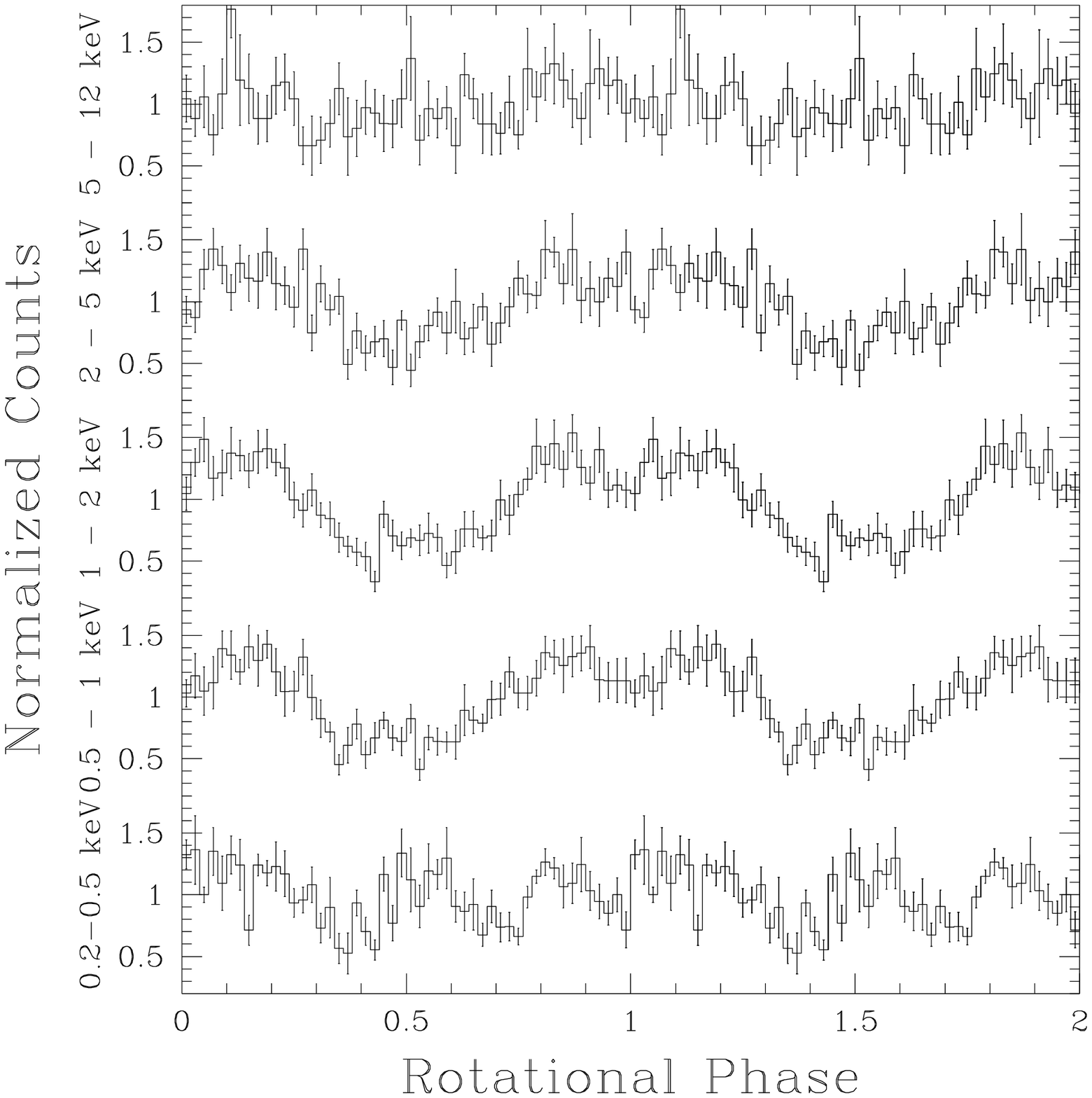}
\epsfxsize=8.cm\epsfysize=8cm\epsfbox{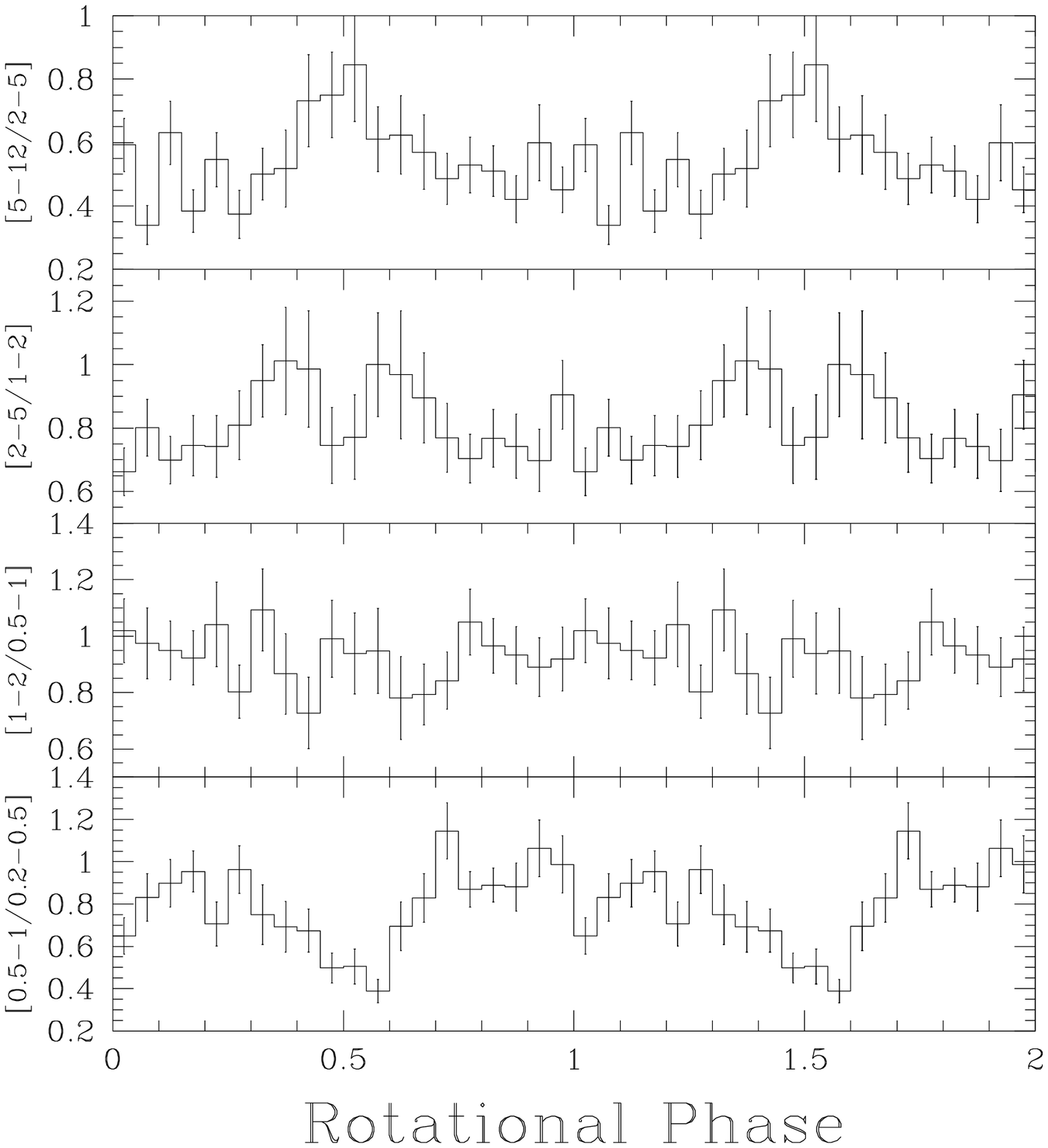}}
\caption{ {\em Left:} EPIC-PN folded light curves in selected energy bands
at the 863.3\,s period using the ephemeris quoted in the text. {\em 
Right:} The EPIC hardness ratios show complex energy dependence of the
pulse.}\label{fig3}
    \end{figure*}

\section{Timing analysis}

\subsection{The  X-ray light curves and power spectra}

A search for variability was performed extracting X-ray light curves from
all available channels of EPIC PN and MOS cameras, i.e. between  
0.2--15\,keV range with a 50\,s 
time resolution (see Fig.~\ref{fig1}).  A periodic modulation is apparent and 
confirmed by a Fourier analysis on the light curve, shown 
in Fig.~\ref{fig2} for the PN camera. 
 Strong power is 
found at 100\,day$^{-1}$ while weaker peaks are detected at its second 
harmonic and at $\sim$92\,day$^{-1}$. The former corresponds to the 
optical 
pulse period found by Burwitz et al. (\cite{Burwitzetal}) and the latter 
to the beat (the orbital sideband $\omega - \Omega$) which was detected 
only at optical wavelengths. No 
sign of low frequency variability is detected in the whole 0.2--15\,keV 
EPIC band.
We then performed a  
sinusoidal fit to the EPIC light curves, including four sinusoids 
accounting for the  fundamental, the first and second harmonic and the 
beat periods. This gives for the PN light curve
$\omega$=100.08$\pm$0.16\,day$^{-1}$ and $\omega - 
\Omega$=92.37$\pm$0.39\,day$^{-1}$, while for the MOS  we
find $\omega$=99.86$\pm$0.24\,day$^{-1}$ and $\omega 
-\Omega$=93.74$\pm$0.60\,day$^{-1}$. Because of the higher fit accuracy of 
the PN 
data, we  then adopt for the spin period: $\rm 
P_{\omega}$=863.3$\pm$1.4\,s.
The time of maximum of the X-ray spin pulse is then: 
 $\rm HJD_{max}$=2\,453239.5648(1)+0.00999(2)\,E. The beat frequency 
implies an orbital period of 
$\rm P_{\Omega}$=3.13$\pm$0.17\,hr, broadly consistent with the optical 
determination by Burwitz et al. (\cite{Burwitzetal}). The amplitude of 
the beat variability is relatively low $\sim 9\%$. A Fourier analysis 
has been also performed on light curves extracted with the same 50\,s bin 
time in selected energy bands, 0.2--0.5\,keV, 0.5--1\,keV, 1--2\,keV, 
2--5\,keV and 5--12\,keV as reported in Fig.~\ref{fig2}.
A different behaviour between soft and hard 
ranges is observed, the second harmonic dominating the softest 
range (0.2--0.5\,keV) with indication of a low frequency periodicity at
9.2\,day$^{-1}$, while between 0.5--5\,keV the spin frequency dominates. 
No variability is detected above 5\,keV.

   \begin{figure}[h!]
   \centering
\includegraphics[height=7.cm, width=8.cm,angle=0]{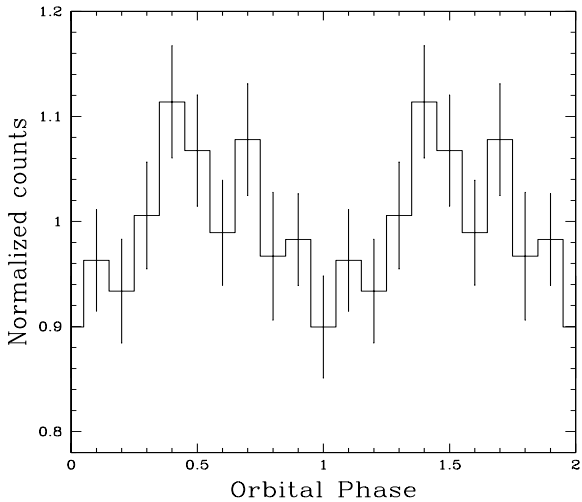}
\caption{The 0.2--0.5\,keV EPIC PN folded light curve at the 3.55\,hr 
orbital period.}\label{fig4}
    \end{figure}

\noindent 
Using the determined spin ephemeris we folded the light curves in 
the different energy ranges as shown 
in Fig.~\ref{fig3}, as well as the
hardness ratios defined as the ratio of countrates in [5--12\,keV] and
[2--5\,keV] ranges, in [2--5\,keV] and [1--2\,keV], in the
 [1--2\,keV] and [0.5--1\,keV]  ranges, in the  
[0.5--1\,keV] and [0.2--0.5\,keV] ranges.  
While above 5\,keV no significant modulation 
is detected, a sawtooth-like spin pulse is observed in the 0.5--5\,keV 
range, with fractional amplitudes (half-amplitude) of 
27$\pm2\%$ 
[2--5\,keV], 37$\pm2\%$ 
[1--2\,keV] and 33$\pm2\%$ [0.5--1\,keV]. A dip 
centred on the maximum of the pulsation more pronounced between  
0.5-2\,keV is observed.  This feature was also observed in V709\,Cas
(Norton et al. \cite{norton99}).
From the hardness ratios the pulse hardens at spin minimum between 
1--12\,keV, while it shows no energy dependence between 0.5--2\,keV. Below 
0.5\,keV and as indicated by the power spectra the spin modulation shows 
instead approximately  three maxima, produced by the dip centred on the 
primary maximum seen in  the harder  bands and an additional maximum 
appearing at the minimum of the hard band
pulse. The appearance of this maximum  is clearly seen in the hardness 
ratios where an  antiphase behaviour is observed with respect to the hard 
bands.

\begin{figure}[h!]
   \centering
\includegraphics[height=9.cm, width=8.cm,angle=0]{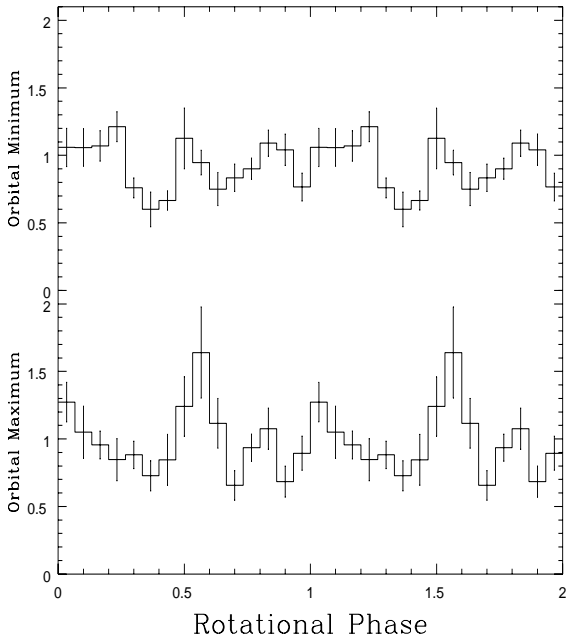}
\caption{ The 0.2--0.5\,keV spin light curves extracted  at orbital 
maximum ({\em lower panel}) and orbital minimum ({\em upper panel}).}\label{fig5}
    \end{figure}

\noindent The 
length of the EPIC coverage is  about twice the 3.5\,hr 
orbital period detected in the optical (Burwitz et al. 
\cite{Burwitzetal}) and ROSAT data (Burwitz \& Reinsch \cite{Burwitz01}).
A sinusoidal fit to the PN light curve in the 0.2--0.5\,keV 
band prewhitened from the high frequency spin 
variability gives an orbital period of 3.55$\pm$0.56hr, thus confirming 
an orbital modulation in the soft X-ray band.  The folded 
orbital light curve, reported in Fig.~\ref{fig4}, shows a  
 modulation with fractional amplitude of $\sim$10$\%$,  similar in shape  but much weaker 
than that found 
in the  ROSAT HRI data  (Burwitz \& Reinsch \cite{Burwitz01}). 
We further explored the orbital dependence of the soft X-ray emission and
in particular the dependence of the soft X-ray spin pulse with the orbital
cycle. We then extracted the spin light curves in the 0.2--0.5\,keV range 
at maximum and minimum of 
the orbital modulation, i.e. between $\rm \phi_{\Omega}$=0.4$\div$ 0.7 and   
$\rm \phi_{\Omega}$=\,-0.2$\div$ +0.2 (a finer binning does not provide 
good statistics of light curves). Fig.~\ref{fig5} shows 
that the pulse profile changes with the orbital phase, being similar to 
the orbital phase--average spin pulse profile at orbital minimum whilst at 
orbital maximum, the spin pulse shows a peak at $\phi_{\omega}\sim$0.5, 
i.e. at  
the minimum of the hard X-ray spin pulse. This behaviour implies that the 
orbital  modulation in the soft X-rays is dominated by a component which 
is anti--phased and hence not linked to the hard X-ray emission. This 
also explains the lack of  an orbital periodicity at high energies.

\subsection{The UV and optical light curves}

\begin{figure}[h!]
   \centering
\includegraphics[height=9.cm, width=8.cm,angle=0]{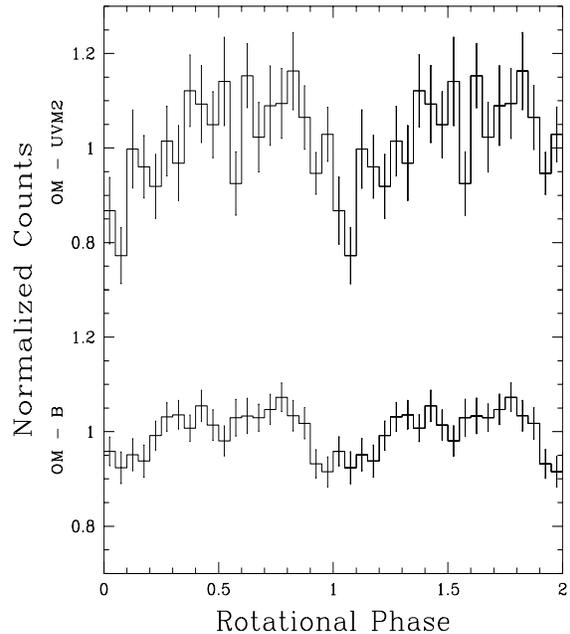}
\caption{ The UV 2000--2800\,AA ({\em upper panel}) and the B band 
spin folded light curves showing a strong colour dependence of 
pulsation.}\label{fig6}
    \end{figure}

The UVM2 and B band light curves shown in Fig.~\ref{fig1} reveal a short 
term modulation, but given the short time coverage of the OM observations 
in  the two bands we limit ourselves to fold the light curves at the 
determined WD spin  period (Fig.~\ref{fig6}). These show an amplitude  
modulation of 
15$\%$ and 6$\%$ in the UV and optical band respectively, suggesting a 
hot component responsible for the pulsation at these wavelengths. The B 
band pulse has similar amplitude  to  that observed by Burwitz et al. 
(\cite{Burwitzetal}). The advantage of simultaneity allows us to compare
X-ray and UV/optical rotational modulations. 
  UU\,Col, in contrast to many IPs, shows an antiphased 
behaviour, with  the minimum of UV and optical pulses  
being centred  on the hard X-ray maximum. The UV and optical rotational  
light curves 
also  do not closely resemble the soft (0.2--0.5\,keV) pulse 
implying that  the latter is composed  of more than one component. 
However the 
maximum of the UV/optical spin pulse is very broad and centred at 
$\phi_{\omega}$=0.5  where the soft band also shows a maximum. 
This indicates that at least part of the soft X-rays are produced 
in a region somewhat linked to that producing the UV and optical 
pulsations. The wider phase range of the UV/optical pulse however
implies a wider emitting area.

\section{The complex X-ray spectrum of UU\,Col}

\begin{figure}
   \centering
\includegraphics[height=8.cm,width=8.cm,angle=-90]{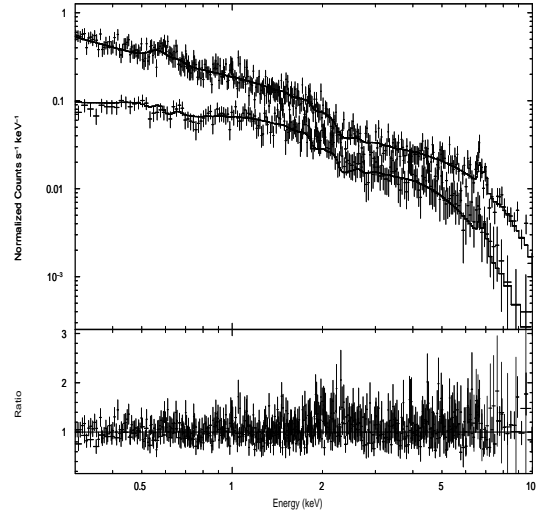}
\caption{The EPIC PN (top) and combined MOS (bottom) spectra fitted 
simultaneously with a composite model consisting of two {\em 
MEKALs} with temperatures kT$_{1}$=11\,keV and kT$_{2}$=0.18\,keV, a 
black-body with kT$_{\rm BB}$=50\,eV a partial absorber with column 
density $\rm N_H=1.0\times
10^{23}$~cm$^{-2}$ and covering fraction 50$\%$.  The bottom panel shows the
ratio between observed and model spectra.}\label{fig7}
    \end{figure}

Identification of  spectral components was performed on the EPIC 
PN and the combined MOS averaged spectra using the XSPEC 
package. Due to 
calibration accuracy issues the spectra were analysed between 0.3 and 
10\,keV (Fig.~\ref{fig7}). 
A simple model consisting of an
optically thin plasma {\sc mekal} (($\rm kT_{MK}$=70\,keV) plus a 
black-body ($\rm kT_{BB}$=80\,eV) does not satisfactorily 
($\chi^2_{\nu}$=1.10) reproduce the  complex spectrum of UU\,Col. A 
 dense ($\rm N_H = 8.3\times10^{22}\,cm^{-2}$) partial (37$\%$) 
covering  absorber is required ($\chi^2_{\nu}$=1.04) and 
significant at 99$\%$ (F-test), which lowers the  
temperature of the optically thin component to 14\,keV. The fit further 
improves ($\chi^2_{\nu}$=1.00)
when the metal abundance of the optically thin component is left  free to 
vary giving $\rm A_Z$=0.41$^{+0.16}_{-0.15}$. Here we note that the 
derived metal abundance is relative to Anders \& Grevesse 
(\cite{andersgrevesse}) 
solar abundances. When adopting cosmic 
abundances as derived for the interstellar medium by Wilms et al. 
(\cite{wilms}), we
find that $\rm A_Z$=0.69$^{+0.31}_{-0.30}$.  This latter gives a slightly 
larger hydrogen column density of the partial absorber ($\rm N_H = 
1.5\times10^{23}\,cm^{-2}$) but still consistent within errors with the 
previous  determination.   
An excess of counts around 0.6\,keV is however still present indicating 
the presence of low temperatures in the post-shock region. Using 
a multi--temperature model ({\sc cemekl}), where the emission measure 
varies with the temperature as $\rm T/T_{max}^{\alpha}$, 
gives instead a worse fit ($\chi^2_{\nu}$= 1.10). The best fit is then 
found adding  another  {\sc 
mekal} component with  kT=0.18\,keV ($\chi^2_{\nu}$=0.906) 
(Model\,A in Table~\ref{spectra}). This also  
lowers the black-body temperature to 50\,eV, which is more typical
for soft X-ray IPs (de Martino et al. \cite{demartino04}). A 
similar fit quality (Model\,B) is obtained by substituting  the hot 
optically thin component with a multi--temperature plasma, 
but many parameters, including an unphysical value for the power law 
index, are unconstrained. 
An upper limit to the equivalent width of the neutral iron line at 
6.4\,keV  is 76\,eV. The flux in the 0.2--10\,keV range is 
$\rm 1.84\pm0.02\,\times10^{-12}\,erg\,cm^{-2}\,s^{-1}$.

   \begin{table*}[t!]
      \caption{Spectral parameters as derived from  
fitting simultaneously the EPIC PN and MOS phase--averaged spectra
for the two best fit models discussed in the text. Quoted errors refer to 
90$\%$confidence level for the parameter of  interest.}
         \label{spectra}
     \centering
\begin{tabular}{ c c c c c c c c c c c c}
            \hline \hline
            \noalign{\smallskip}
 & \multicolumn{2}{c}{Partial Absorber } & \multicolumn{2}{c}{Black-body} 
 & \multicolumn{3}{c}{{\sc mekal}  } & \multicolumn{3}{c}{{\sc 
mekal}$^{a}$ } & $\chi^2_{\nu}$ ($\chi^2$/d.o.f.) \cr
            \noalign{\smallskip}
 & $\rm N_H^{b}$   & $\rm Cov_{F}^{c}$ & $\rm kT_{BB}$ & $\rm 
C_{BB}^{d}$ & 
$\rm A_Z^{e}$ & $\rm kT_{1}$ & $\rm C_{1}^{f}$ & $\rm 
kT_{2}^{g}$ &
$\rm C_{2}^{f}$ & $\alpha^{h}$ &  \cr
            \noalign{\smallskip}
& ($10^{23}$~cm$^{-2}$) & & (eV) & $(10^{-5})$ &  & (keV) & $(10^{-4})$ & 
(keV) & $(10^{-4})$ &  \cr 
            \hline
            \noalign{\smallskip}
{\bf A} & 1.0$^{+0.4}_{-0.3}$ & 0.51$^{+0.07}_{-0.09}$ & 
49.7$^{+5.6}_{-2.9}$ & 1.3$^{+0.5}_{-0.3}$ & 0.39$\pm$0.16 &
0.18$\pm$0.02 &  2.4$^{+1.1}_{-0.9}$ & 11$^{+6}_{-2}$ & 
19.04$^{+0.38}_{-1.98}$ & & 0.906 (522/576) \cr
            \noalign{\smallskip}
{\bf B } & 0.83$^{+0.51}_{-0.32}$ & 0.43$^{+0.15}_{-0.17}$ & 
51.2$\pm$6.0 &  1.04$^{+0.63}_{-0.28}$ & 0.53$^{+0.71}_{-0.26}$ & 
0.18$^{+0.02}_{-0.01}$ & 1.6$^{+2.0}_{-1.0}$ & 27$^{+\infty}_{-17}$ & 
53.48$^{+68.16}_{-19.75}$ & 1.6$^{+2.9}_{-0.40}$ & 0.905 (520/575) \cr
            \noalign{\smallskip}
            \hline
            \hline
\end{tabular}
~\par
\begin{flushleft}
$^a$: Second optically thin plasma is a {\sc mekal} (Model {\bf A}) or 
{\sc cemekl} (Model {\bf B}).\par
$^b$: Column density of the partial absorber.\par
$^c$: Covering fraction of partial absorber.\par
$^d$: Normalization constant of black-body ({\sc bbody}) defined 
as $\rm  L_{39}\,d_{10}^{-2}$, where $\rm L_{39}$ is the luminosity in 
units of  10$^{39}\,erg\,s^{-1}$ and $\rm ,d_{10}$ the distance in units 
of 10\,kpc.\par
 $^e$: Metal abundance in units of the cosmic value (Anders \& Grevesse
\cite{andersgrevesse}) linked for the two thin plasma emissions.\par
$^f$: Normalization constant of {\sc mekal} or {\sc cemekl } model 
defined as $\rm E.M.\times10^{-14}\,(4\,\pi\,D^2)^{-1}$, where E.M. is the 
emission  measure in $\rm cm^{-3}$ and D is the distance in cm.  \par
$^g$: Maximum temperature in {\sc cemekl}.\par
$^h$: Power law  index in {\sc cemekl}.\par  
\end{flushleft}
\end{table*}

\begin{figure}
   \centering
\includegraphics[height=8.cm,width=8.cm,angle=0]{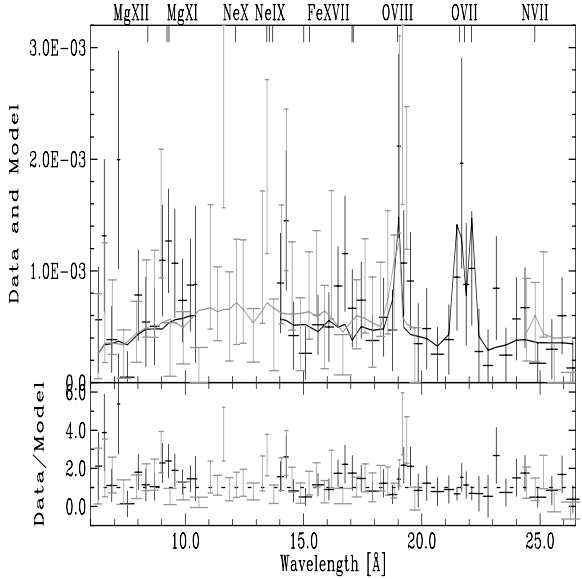}
\caption{The RGS spectra (RGS1 in black and RGS2 in gray) along 
with Model\,A  derived from the EPIC 
spectral analysis. The positions of lines expected to be strong are also 
reported. The ratio 
between model and observed spectrum is shown in the lower 
panel.}\label{fig8}
    \end{figure}

\noindent The excess in the 0.6\,keV region can also be reproduced  by 
two gaussians centred  at 0.57\,keV and 0.65\,keV ($\chi^2_{\nu}$= 0.893) 
implying that  {\sc O\,VII} (21.9\,$\AA$) He-like triplet and {\sc 
O\,VIII} (19.1\,$\AA$) Ly$_{\alpha}$ line are present. 
Though UU\,Col is quite faint in the RGS spectra, we can safely detect 
both lines at $\rm E_{\sc O\,VII}=0.570^{+0.012}_{-0.021}$\,keV 
at a flux $\rm 1.5^{+3.8}_{-1.2}\times 10^{-5}\,photons\,cm^{-2}\,s^{-1}$ 
in the RGS1  and at $\rm E_{\sc 
O\,VIII}=0.648^{+0.002}_{-0.006}$\,keV 
at a flux $\rm 1.2^{+0.8}_{-0.6}\times 10^{-5}\,photons\,cm^{-2}\,s^{-1}$ 
in  both RGS.   The RGS spectra when fitted with Model\,A give 
$\chi^2_{\nu}$= 1.3 for 46 d.o.f. 
The large width ($\sim$6 eV) of the {\sc  O\,VII} line suggests 
the presence  of multiple components. An enlargment of the Oxygen region  
is shown in  Fig.~\ref{fig8} rebinning the spectra to have a minimum 
of 9 counts per bin.  We then fitted the {\sc  O\,VII} line 
spectral region with two  gaussians and a power law which gives
for these components central 
energies of 0.572\,keV and 0.561\,keV.  The former is broader 
($\sigma\sim$ 
2.4\,eV) and stronger ($\sim$28 times) than the latter.
The narrow line  at 
22.10\,$\AA$ can be ascribed to the forbidden {\em (f)} line component 
while  the broad feature is probably a blend of 
the resonance  {\em (r)} (21.603\,$\AA$)  and the intercombination {\em 
(i)} (21.796\,$\AA$) lines.  
The relatively large strength of He-like Oxygen is consistent with the 
presence of a low temperature optically thin component. This aspect will 
be further discussed in sect.\,5.

In order to identify spectral changes with the spin phase, we  extracted 
the EPIC PN spectra (the inclusion of MOS 
spectra does not improve substantially the statistics) at pulse maximum 
and minimum 
identified on the total 0.2--15\,keV  light curve,
i.e. between 
$\phi_{\omega}=-0.2\div$ +0.2 and $\phi_{\omega}= 0.3\div$ 0.65 
respectively. 
The spectral fitting was performed using Model\,A in Table~\ref{spectra}
fixing the metal abundance to the value found for the average spectrum, 
but many  of the resulting parameters were unconstrained. Since no 
spectral 
variability is detected above 5\,keV (see Fig.~\ref{fig3}), we then kept  
both  the normalization 
and  the  temperature of the hot plasma fixed at the values found 
for the average spectrum.
From Table~\ref{spin_spectra} no substantial change in the parameters is 
found except for the partial absorber implying that above 0.5\,keV this 
component is responsible for the modulation. On the other hand 
we are unable to identify the components responsible for the 
soft X-ray pulsation.

\section{Discussion}

Our {\em XMM-Newton} observation of UU\,Col has revealed new properties 
of this poorly studied magnetic system.

   \begin{table}
      \caption{Spectral parameters as derived from  
fitting the EPIC PN spectrum at maximum and minimum of spin pulsation 
using Model\,A in Table~\ref{spectra}.}
         \label{spin_spectra}
     \centering
\begin{tabular}{ c c c }
            \hline \hline
            \noalign{\smallskip}
 & Pulse Maximum & Pulse Minimum \cr
            \noalign{\smallskip}
            \hline
            \noalign{\smallskip}
$\rm F_{2-10}$ & 1.90$\pm$0.02 &  1.74$\pm$0.03 \cr 
$[10^{-12}\,erg\,cm^{-2}\,s^{-1}]$  &    & \cr
            \noalign{\smallskip}
            \hline
            \noalign{\smallskip}
$\rm N_H\,[10^{23}$~cm$^{-2}]$   & 1.01$^{+0.29}_{-0.34}$ & 1.33$^{+0.41}_{-0.27}$    \cr
$\rm Cov_{F}$ & 0.49$^{+0.03}_{-0.02}$ & 0.56$\pm$0.03 \cr
$\rm kT_{BB}$\,[keV] &  50$^{+11}_{-10}$ & 51$^{+\infty}_{-10}$ \cr
$\rm C_{BB}\,[10^{-5}]$ & 1.25$^{+1.03}_{-0.75}$ & 1.13$^{+1.07}_{-\infty}$ \cr
$\rm kT_{1}$\,[keV] & 0.16$^{+0.02}_{-0.04}$ & 0.17$^{+0.04}_{-0.03}$ \cr
$\rm C_{1}\,[10^{-4}]$ & 3.4$^{+4.2}_{-1.4}$ & 3.0$^{+11.6}_{-0.9}$ \cr 
$\rm A_Z^{a}$ & 0.39 & 0.39 \cr
$\rm kT_{2}^{a}$\,[keV]  & 11 & 11 \cr
$\rm C_{2}^{a}\,[10^{-4}]$ & 19.04 & 19.04 \cr
$\chi^2_{\nu}$ ($\chi^2$/d.o.f.) & 0.90 (127/141) & 1.01 (111/110) \cr
            \noalign{\smallskip}
            \hline
            \hline
\end{tabular}
~\par
\begin{flushleft}
$^a$: Parameter fixed at the value found for the average spectrum.\par
\end{flushleft}
\end{table}

\subsection{The pulsation properties}

The X-ray variability is dominated by the 863.3\,s pulsation ascribed
to the WD rotational period by Burwitz et al. \cite{Burwitzetal}. 
The detection for the first time of a weak (9$\%$) 935.4\,s X-ray 
periodicity, consistent with the $\omega - \Omega$ orbital sideband,  
indicates that two accretion modes exist (see Hellier \cite{hellier}; 
Norton et al. \cite{Nortonetal}). In fact, the presence of a 
dominant (full amplitude $\sim75\%$) spin pulsation implies that accretion 
occurs mainly via a disc 
because the material circulating in the disc looses memory of the orbital 
motion; but an additional beat ($\sim 20\%$) variability indicates 
that part of the accreting material also flows directly towards the WD 
without passing through the disc (disc-overflow).  A similar behaviour was 
observed in a few other IPs like TX\,Col (Norton et al. \cite{norton97}) and 
FO\,Aqr (Beardmore et al. \cite{beardmore98}). 
 The power spectrum also shows substantial power at the second 
harmonic of the spin frequency (3$\omega$). A similar feature was also 
observed in the power spectrum of  V709\,Cas (Norton et al. 
\cite{norton99}; de Martino et al. \cite{demartino01}) and interpreted as 
the contribution of two 
asymmetric accreting poles. However,  
energy resolved power spectra show that in UU\,Col the 3$\omega$ frequency 
dominates below 0.5\,keV implying that the accretion pattern and its 
associated emission properties are very complex. Furthermore
the presence of an orbital modulation only in the soft X-ray 
range imply that the  region responsible is not 
strictly related to that producing the hard X-ray emission. This aspect 
is further explored below. 

The complex energy dependence of the spin pulse suggests 
the contribution of multiple components. 
 No modulation is detected above 5\,keV indicating no influence from 
the hottest post--shock regions on the pulsation, but between 
0.5--5\,keV  the spin is  likely due to  photo-electric absorption. This 
behaviour is  observed in most IPs  (Norton \&  Watson \cite{Norton89}; 
de Martino et al. \cite{demartino01}, \cite{demartino04}). 
The hardening at spin minimum and the identification 
of a dense ($\rm 10^{23}\,cm^{-2}$) absorber covering about 50$\%$  
of the X-ray  source, is consistent  with the standard accretion curtain scenario 
of IPs (Rosen et al. \cite{rosen}), 
where the material from the accretion disc is captured by the 
magnetic field lines flowing towards the WD poles in an arc--shaped 
curtain. The complexity and inhomogeneity of the absorber in IPs was first 
recognized by  Mukai et al. (\cite{mukai}).
The appearance of an additional maximum in the soft 
0.2--0.5\,keV band at the minimum of the hard pulse indicates that besides 
the main accreting pole, which is active in both soft and 
hard X-ray ranges, there 
is a  substantial soft X-ray contribution from the secondary pole.
It therefore appears that the soft X-rays comprise of contributions 
from both the primary and secondary poles. This behaviour, though with 
its 
individual differences,  is also seen in 
the soft IP V405\,Aur (Evans \& Hellier 
\cite{evanshellier}) but not in the other IPs of this small 
group of soft X-ray systems (de Martino et al. \cite{demartino04}). In 
V405\,Aur the soft spin pulse is double-peaked,  while in UU\,Col 
the light curve has three maxima indicating that the accretion geometry
in the latter is more complex.
The  presence of an orbital modulation only in the soft X-rays, not 
detected in V405\,Aur,  
  and the fact that the spin pulse at orbital maximum 
reveals a strong  contribution from the secondary pole might favour the 
interpretation that material flowing directly from the stream impacts 
preferentially onto this pole. Also, the fact that the spin light curve 
has three peaks might suggest an asymmetry in the overflow impact regions.
Unfortunately from our spectral analysis at maximum and at minimum of the 
spin pulsation we cannot establish
whether the secondary pole is dominated by the black--body soft X-ray 
emission or the optically thin cool component.
However we have found  that the UV and optical spin pulsations are 
anti--phased  with respect to the hard X-ray pulse and that they show a 
broad  maximum and strong colour dependence. This implies that 
UV/optical rotational modulation does not originate in the accretion 
curtain  above the main accreting pole but it arises from the heating of a 
large area of the WD surface 
at the secondary pole. From the amplitudes of UV and B band spin 
modulations ($15\%$ and 6$\%$ respectively) we derive a colour 
temperature  of $\sim$31000\,K assuming a 
black--body emission of the region producing the UV/optical spin modulated 
flux. Furthermore the pulsed fluxes at the 
effective 
wavelengths of the UVM2 and B filters imply for a WD with Log\,g=8.0 and 
T$_{\rm eff}$=30000\,K, a radius of $\rm \sim 
7\times10^{7}\,D_{100pc}\,cm$, where $D_{100pc}$ is the distance in units 
of 100\,pc.
Based on the K band surface brightness Burwitz et al. 
(\cite{Burwitzetal}) 
estimate a lower limit of 740\,pc for the distance of UU\,Col, which would 
imply $\rm R_{spot} \geq  5.2\times10^{8}$\,cm, and hence a relatively 
large spot of the WD surface.
 It may therefore occur that the two-mode 
accretion onto the WD involves both poles with the secondary 
mainly emitting radiation at lower energies.  

\subsection{The accretion flow properties }      

The X-ray spectral analysis reveals the presence of multiple components: a 
hot  optically thin plasma  at 11\,keV which is  visible  throughout  
the spin  cycle. The emission measure derived for this component is $\rm 
EM_{hot}=2.3\times10^{53}\,D_{100pc}^2\,cm^{-3}$ and its bolometric 
luminosity is $\rm 6.4\times 10^{30}\,D_{100pc}^2\,erg\,s^{-1}$. A 
multi--temperature 
power--law structure of the post--shock region seems not to be required 
though  a cooler thin emission at 0.18\,keV, is clearly present. This low
temperature plasma is also detected in V405\,Aur (Evans \& 
Hellier \cite{evanshellier}). We derive for this component an emission 
measure  $\rm EM_{cool}=2.9\times^{52}\,D_{100pc}^2\,cm^{-3}$  and a 
bolometric luminosity of $\rm 1.1\times 
10^{30}\,D_{100pc}^2\,erg\,s^{-1}$,  hence 
$\sim$8 and $\sim$ 6
times smaller than those of the hot optically thin emission, respectively.  
It is likely  that this component originates  much closer to the WD 
surface than the hotter region. 
The strong {\sc OVIII} and {\sc OVII} lines detected in the RGS spectra, 
which  particularly map the low temperature plasma conditions, 
are well accounted by for the cool thin plasma component. The  flux 
ratio of these lines indeed indicates a temperature of 
$0.2^{+0.1}_{-0.03}$\,keV. The {\sc OVII} He-like features 
tentatively identified as the resonant and forbidden components give a 
ratio {\em r}/{\em f}$\sim$28 suggesting a collision-dominated plasma 
(Porquet 
et al. \cite{porquet01}). We are unable to derive an estimate of the 
density because of the weakness of the features in the RGS spectra.

The X-ray spectrum also shows a black--body soft 
X-ray  component at 50\,eV, similar to that found for 
PQ\,Gem 
and V405\,Aur,  the other two bright soft 
X-ray IPs (de Martino et al. \cite{demartino04}; Evans \& Hellier 
\cite{evanshellier}).  Its bolometric 
 luminosity is $\rm 1.3\times10^{30}\,D_{100pc}^2\,erg\,s^{-1}$. The 
emitting area of this component is $\rm a_{BB}=2.0\times 
10^{11}\,D_{100pc}^2\,cm^2$. 
At the minimum distance of 740\,pc, this gives  $\rm r_{BB}\sim 
2\times10^{6}\,cm$ much 
smaller than the radius of the UV/optical emitting region.  
The ratio of bolometric fluxes between the soft X-ray 
black--body and hard X-ray components is only 0.20 and hence  lower 
than that estimated by Burwitz et al. (\cite{Burwitzetal}) who assumed  
a simple 
model consisting of a black--body at 25\,eV and a thermal Bremsstrahlung 
at 20\,keV. However we note that using their same simple model we derive a 
ratio $\rm  F_{BB}/F_{th.}\sim$ 13. It is therefore clear that previous 
determination of the energy balance is  subject to strong revisions 
due to proper determination of the spectral components.

We have seen that the secondary pole mainly radiates at low energies and 
it is responsible for the orbital modulation in the soft X-rays. This 
might suggest that this pole is fed predominantly by material not circulating 
in  the disc.  To estimate the accretion luminosity we then take into
account the different spectral components. Hence, 
assuming that the accretion luminosity is emitted in the hard, soft 
X-rays and UV/optical wavelengths, $\rm L_{acc}= L_{BB} + 
L_{th.} + L_{UV/Opt} =9.5\times10^{30}\,D_{100pc}^{2}\,erg\,s^{-1}$, where 
we include both optically thin components in the bolometric flux 
computation,  we derive an accretion rate of 
$\rm 1.5\times10^{-12}\,D_{100pc}^{2}\,M_{\odot}\,yr^{-1}$, for  
a WD mass of 0.6\,M$_{\odot}$. At the 
minimum distance of 740\,pc the mass accretion rate is much smaller than 
the secular rate predicted for a CV with a 3.5\,hr orbital period 
($\rm \dot M = 2.0\times10^{-11}\,P_{\Omega,hr}^{3.7}\,M_{\odot}\,yr^{-1}$ 
(see Warner \cite{warner95})),  
unless the distance is extremely large ($\sim$ 3.5\,kpc). It therefore seems 
that UU\,Col is a soft X-ray IP with a low mass accretion rate. 
The lower limit to the distance and the high latitude would put this 
system well above the galactic disc and about 2-3 scale heights of the CV 
population. This makes it unlikely that UU\,Col belongs to the disc 
population. If it is a rare case of a halo CV this might 
explain its very low mass accretion rate. 

The ratio of spin-to-orbit periods is $\rm P_{\omega}/P_{\Omega}$=0.07 
close to 0.1, which is the typical value of
period ratios of IPs. Though we have detected that accretion also occurs
directly from the stream, the bulk of material is accreted via a disc.
We then investigate the spin equilibrium state evaluating first the 
corotation radius, defined as the radius at which 
the magnetic field rotates at the Keplerian frequency: $\rm 
R_{co}=(G\,M_{WD}\,P_{\omega}^{2}/4\,\pi^2)^{1/3}$. For a rotational 
period 
of 863\,s and  $\rm M_{WD}=0.6\,M_{\odot}$,  $\rm 
R_{co}=1.2\times10^{10}\,cm$. The condition for accretion requires 
$\rm R_{mag} \leq R_{co}$ with $\rm R_{mag} = 5.5\times
10^{8}\,(M_{WD}/M_{\odot})^{1/7}\,R_{9}^{-2/7}\,L_{33}^{-2/7}\,\mu_{30}^{4/7}$\,cm,
where $\rm R_{9}$ is the WD radius in units of $10^{9}$\,cm, $\rm
L_{33}$ is the luminosity in units of $\rm 10^{33}\,erg\,s^{-1}$ and 
$\mu_{30}$ is the magnetic moment in units of $\rm 10^{30}$\,G\,cm$^3$.
Using our mass accretion rate determination we then find  
$\rm \mu \leq 2\times10^{31}\,D_{100pc}\,G\,cm^3$. Norton et al. 
(\cite{norton04}) derive a magnetic moment of 
$\rm 6.5\times10^{32}\,G\,cm^3$ 
for UU\,Col applying their magnetic accretion model and assuming it is in 
spin equilibrium. Their estimate of the magnetic moment is about 4 
times larger than our  upper limit for the minimum distance of 740\,pc, 
but we 
note that they assume  an accretion rate corresponding to the secular 
value predicted for its  orbital period, which we found is too large for 
UU\,Col.  Hence UU\,Col appears to possess a 
a weak  magnetic field WD accreting at a relatively low rate. The 
condition of 
spin equilibrium is attained when $\rm  
R_{circ}\sim R_{co}$, where $\rm R_{circ}$
is the radius at which  material from the stream  leaving the secondary 
star at the inner Lagrangian point L$_1$ circulates in a Keplerian orbit 
around the WD, defined as 
$\rm R_{circ} \simeq 4\,\pi^2\, P_{\Omega}^{-2}\,b^4\,(G\,M_{WD})^{-1}$, 
with b the distance of L$_1$ from the WD. Assuming a typical mass ratio 
$q=0.5$, we derive $\rm R_{circ} \sim 1.5\times10^{10}$\,cm. Hence UU\,Col 
is close to its equilibrium value. 

\section{Conclusions}

Based on  the  first XMM-Newton observations our X-ray study of 
UU\,Col, complemented by  UV and optical simultaneous photometry,  
has shown new interesting properties of this poorly studied faint soft 
X-ray IP. These can be summarised as follows:

\begin{itemize}

\item The X-ray (0.2--15\,keV) variability is dominated by the 863\,s spin 
periodicity  but a weak variability is also detected at the 935\,s 
beat period. This indicates two accretion modes operating in this system 
with the bulk ($\sim 80\%$) of accretion occurring via a disc and with a 
small fraction ($\sim 20\%$) 
accreting directly from the stream.

\item The spin pulse shows a complex energy dependence  revealing the 
presence of the secondary pole which mainly emits below 0.5\,keV. At 
intermediate energies between 0.5--5\,keV, the spin pulse is likely due 
to photoelectric absorption. 
Above 5\,keV no modulation is observed indicating that the hottest 
post--shock regions of the main accreting 
pole are viewed throughout the spin cycle. 
 
\item We also detected an orbital periodicity below 0.5\,keV without any 
counterpart in the hard X-rays. The spin pulse shape at maximum and 
minimum of the orbital cycle suggests a dominant contribution 
from the secondary pole.

\item The UV and optical pulses show a broad maximum at the minimum of the 
hard X-ray pulsation, revealing that they arise from a large region at the 
WD  surface close to the secondary pole. 

\item The X-ray spectrum of UU\,Col is complex,  
consisting of a soft X-ray black--body component at $\sim$50\,eV and two 
optically thin emissions at 0.2\,keV and 11\,keV, highly absorbed by dense 
($\rm 10^{23}\,cm^{-2}$) material partially (50$\%$) covering the X-ray 
source.  The RGS spectra reveal strong {\sc OVIII}  and {\sc OVII}
lines  confirming the presence of the cooler optically thin emitting 
region. 

\item The ratio of soft-to-hard bolometric fluxes is not high ($\rm
F_{BB}/F_{th}$=0.2) consistent with the prediction of the standard model
of accretion shocks.

\item From the X-ray, UV and optical luminosities we infer a mass 
accretion 
rate of $\rm 
1.5\times10^{-12}\,D_{100pc}^{2}\,M_{\odot}\,yr^{-1}$,  lower than 
the secular value expected for its 3.5\,h orbital period. 

\item The WD in UU\,Col is found to possess a low magnetic moment and 
 to rotate at its equilibrium value. Therefore, UU\,Col appears to be 
different  from the other two well studied soft X-ray bright IPs PQ\,Gem 
and V405\,Aur, which 
instead possess stronger magnetic fields. Hence UU\,Col might not 
evolve into a moderate field strength Polar thus leaving the soft X-ray 
systems a still enigmatic small group of IPs.

\end{itemize}

 \begin{acknowledgements}
DDM acknowledges financial support by the Italian Minister of 
University and Research (MIUR) and  the Italian Space Agency (ASI) under 
contract I/023/05/0.
 \end{acknowledgements}

\end{document}